\documentclass[11pt,preprintnumbers,titlepage,nofootinbib]{revtex4-2}%
\usepackage[latin9]{inputenc}
\usepackage{amsmath}
\usepackage{amssymb}
\usepackage{graphicx}
\usepackage[unicode=true,  bookmarks=false,  breaklinks=false,pdfborder={0 0
1},backref=section,colorlinks=false]{hyperref}
\usepackage{wasysym}
\usepackage{array}
\usepackage{multirow}
\usepackage{wasysym}
\usepackage{wasysym}
\usepackage{amsfonts}
\usepackage{subfigure}
\usepackage{cleveref}
\usepackage{listings}
\usepackage{xcolor}
\usepackage{array}
\usepackage{url}
\usepackage{color}%
\hypersetup{
colorlinks,linkcolor=blue,citecolor=blue,urlcolor=blue}
\makeatletter
\newcommand{\lyxdot}{.}

\@ifundefined{textcolor}{}{
\definecolor{BLACK}{gray}{0}
\definecolor{WHITE}{gray}{1}
\definecolor{RED}{rgb}{1,0,0}
\definecolor{GREEN}{rgb}{0,1,0}
\definecolor{BLUE}{rgb}{0,0,1}
\definecolor{CYAN}{cmyk}{1,0,0,0}
\definecolor{MAGENTA}{cmyk}{0,1,0,0}
\definecolor{YELLOW}{cmyk}{0,0,1,0}
}

\makeatother
\begin{document}
\preprint{CTP-SCU/2025016}
\title{Rotating Black Holes in Einstein-Born-Infeld Theory}
\author{Lang Cheng}
\email{chenglang@stu.scu.edu.cn}
\author{Peng Wang}
\email{pengw@scu.edu.cn}
\affiliation{College of Physics, Sichuan University, Chengdu, 610064, China}

\begin{abstract}
We numerically construct a family of stationary, axisymmetric black hole
solutions in Einstein-Born-Infeld theory, incorporating both electric charge
and rotation. Our results indicate that when nonlinear electromagnetic effects
are weak, rotating BI black holes with fixed spin approach the extremal limit
as the electric charge increases. In contrast, strong nonlinear effects lead
to the termination of solutions at configurations corresponding to naked
singularities. We demonstrate that nonlinear electrodynamics enhances the
gyromagnetic ratio relative to that of Kerr-Newman (KN) black holes.
Additionally, we analyze the Innermost Stable Circular Orbits (ISCOs) and find
that both prograde and retrograde ISCO radii are consistently smaller than
those found in KN black holes.

\end{abstract}
\maketitle
\tableofcontents

{}

{}

\section{Introduction}

Black holes are among the most fascinating and fundamental objects in
gravitational physics, serving as theoretical laboratories for testing
classical and quantum aspects of gravity. In general relativity coupled to
Maxwell electrodynamics, the Kerr-Newman (KN) solution represents the most
general stationary, asymptotically flat black hole with mass, angular
momentum, and electric charge \cite{Newman:1965my}. However, it is widely
believed that Maxwell's linear theory may be inadequate in the strong-field
regime. A natural extension is to consider nonlinear electrodynamics
\cite{Soleng:1995kn,AyonBeato:1998ub}, among which Born-Infeld (BI) theory
stands out as a particularly well-motivated example. Originally introduced to
regularize the infinite self-energy of point charges \cite{Born:1934gh}, BI
theory also emerges in the low-energy limit of string theory and significantly
alters the dynamics of electromagnetic fields at high field strengths
\cite{Fradkin:1985qd,Tseytlin:1986ti,Salazar:1987ap,Wiltshire:1988uq}.

Einstein-Born-Infeld (EBI) theory, which couples BI electrodynamics to
gravity, has received considerable attention in recent years, especially in
the study of static, spherically symmetric solutions. The first such solutions
were obtained in asymptotically flat spacetime \cite{Demianski:1986wx} and
were subsequently extended to include a cosmological constant
\cite{Fernando:2003tz,Dey:2004yt,Cai:2004eh}. These black holes exhibit a
range of distinctive features and have been the focus of extensive research,
including investigations of their thermodynamic properties
\cite{Rasheed:1997ns,Breton:2004qa,Banerjee:2010da,Zou:2013owa,Hendi:2015hoa,Wang:2019kxp,Liang:2019dni,Wang:2018xdz}%
, quasinormal modes \cite{Breton:2017hwe,Lee:2020iau}, and optical appearance
\cite{He:2022opa,Chen:2023trn,Chen:2023knf}, as well as other physical and
phenomenological aspects
\cite{Tao:2017fsy,Gan:2019jac,Peng:2018smy,Wang:2020ohb,Nomura:2020tpc,Yang:2020jno,Falciano:2021kdu,Babar:2021nst,Yang:2022qoc,Ali:2023jox,Tang:2023lmr,Wu:2024fvy,Maier:2024ahy,Zahid:2024nvx,Wu:2024qpf,Ye:2025xvi,Chen:2025cwi}%
.

However, much less is known about their rotating counterparts, despite the
astrophysical importance of black hole spin. In \cite{CiriloLombardo:2004qw},
rotating BI black hole metrics were derived using the Newman-Janis algorithm;
however, these do not correspond to exact solutions of the EBI field
equations. Approximate solutions have also been explored in the slowly
rotating limit \cite{CiriloLombardo:2005qmc}. Owing to the inherent
nonlinearities of the BI action, numerical methods are essential for
constructing rotating black hole solutions. In this work, we aim to construct
such solutions by numerically solving the full set of nonlinear equations of motion.

In this paper, we also examine two key physical properties of rotating BI
black holes: the gyromagnetic ratio and Innermost Stable Circular Orbit
(ISCO). The gyromagnetic ratio $g$ characterizes how the magnetic dipole
moment is induced by the total angular momentum and charge for a given mass.
For fundamental particles like the electron, the Dirac theory predicts $g=2$.
Remarkably, KN black holes in Einstein-Maxwell theory also possess a
gyromagnetic ratio of $g=2$ \cite{Carter:1968rr}, a coincidence that has
historically prompted speculation about deep connections between black holes
and fundamental particles. Since then, numerous studies have investigated the
gyromagnetic ratio of rotating charged black holes beyond the KN solution
\cite{Garfinkle:1990ib,Aliev:2004ec,Ortaggio:2006ng,Aliev:2006tt}. Notably,
departures from $g=2$ can occur in black holes with scalar hair, offering
potential astrophysical signatures that could distinguish among competing
black hole models \cite{Delgado:2016jxq}.

Another key observable is the radius of the ISCO, which marks the inner edge
of accretion disks around black holes. In the KN spacetime, ISCO radii depend
sensitively on both the spin and the charge of the black hole, with distinct
values for prograde and retrograde orbits. These orbits directly influence the
thermal and dynamical properties of the accretion disk and therefore affect
electromagnetic and gravitational wave signals. ISCOs also provide a way to
probe the near-horizon geometry, and deviations from their KN counterparts can
reflect the presence of modified matter fields or corrections to general relativity.

The structure of the paper is as follows. In Sec. \ref{sec:Set-Up}, we
introduce the EBI theory and outline the numerical method used to construct
rotating black hole solutions. Sec. \ref{sec:NR} presents our numerical
results and examines the properties of rotating BI black holes. Finally, our
conclusions are summarized in Sec. \ref{sec:Conclusion}. Throughout this work,
we adopt the convention $G=c=4\pi\varepsilon_{0}=1$.

\section{Set Up}

\label{sec:Set-Up}

In this section, we first introduce the gravitational model coupled to a BI
electromagnetic field. We then describe the numerical method used to solve the
coupled nonlinear partial differential equations, including the specific
ansatz for the metric and gauge fields, as well as the imposed boundary conditions.

\subsection{Einstein-Born-Infeld Theory}

We consider a $\left(  3+1\right)  $ dimensional gravitational model coupled
to a BI electromagnetic field $A_{\mu}$. The action $\mathcal{S}$ is given by
\begin{equation}
\mathcal{S}=\frac{1}{16\pi}\int dx^{4}\sqrt{-g}\left[  R+4\mathcal{L}\left(
s,p\right)  \right]  , \label{eq:NLEDAction}%
\end{equation}
where $R$ is the Ricci scalar, and the BI Lagrangian $\mathcal{L}\left(
s,p\right)  $ is defined as
\begin{equation}
\mathcal{L}\left(  s,p\right)  =\frac{1}{a}\left(  1-\sqrt{1-2as-a^{2}p^{2}%
}\right)  \text{.}%
\end{equation}
Here, $s$ and $p$ are two independent scalars constructed from the field
strength tensor $F_{\mu\nu}=\partial_{\mu}A_{\nu}-\partial_{\nu}A_{\mu}$,
without involving any of its derivatives,
\begin{equation}
s=-\frac{1}{4}F^{\mu\nu}F_{\mu\nu}\text{ and }p=-\frac{1}{8}\epsilon^{\mu
\nu\rho\sigma}F_{\mu\nu}F_{\rho\sigma}\text{,}%
\end{equation}
where $\epsilon^{\mu\nu\rho\sigma}\equiv-\left[  \mu\text{ }\nu\text{ }%
\rho\text{ }\sigma\right]  /\sqrt{-g}$ is a totally antisymmetric Lorentz
tensor, and $\left[  \mu\text{ }\nu\text{ }\rho\text{ }\sigma\right]  $
denotes the permutation symbol. The coupling parameter $a$ is related to the
string tension $\alpha^{\prime}$ as $a=\left(  2\pi\alpha^{\prime}\right)
^{2}$.

The equations of motion are obtained by varying the action $\left(
\ref{eq:NLEDAction}\right)  $ with respect to $g_{\mu\nu}$ and $A_{\mu}$,
yielding
\begin{align}
R_{\mu\nu}-\frac{1}{2}Rg_{\mu\nu}  &  =8\pi T_{\mu\nu}\text{,}\nonumber\\
\nabla_{\mu}\left[  \frac{\partial\mathcal{L}\left(  s,p\right)  }{\partial
s}F^{\mu\nu}+\frac{1}{2}\frac{\partial\mathcal{L}\left(  s,p\right)
}{\partial p}\epsilon^{\mu\nu\rho\sigma}F_{\rho\sigma}\right]   &  =0\text{,}
\label{eq:NLEDEOM}%
\end{align}
where the energy-momentum tensor $T_{\mu\nu}$ is given by
\begin{equation}
T_{\mu\nu}=\frac{1}{4\pi}g_{\mu\nu}\left[  \mathcal{L}\left(  s,p\right)
-p\frac{\partial\mathcal{L}\left(  s,p\right)  }{\partial p}\right]  +\frac
{1}{4\pi}\frac{\partial\mathcal{L}\left(  s,p\right)  }{\partial s}F_{\mu\rho
}F_{\nu}^{\;\rho}\text{.}%
\end{equation}
In the limit $a\rightarrow0$, the BI Lagrangian $\mathcal{L}\left(
s,p\right)  $ reduces to the standard Maxwell form, and the corresponding
black hole solutions recover the KN family.

The equations of motion $\left(  \ref{eq:NLEDEOM}\right)  $ admit static,
spherically symmetric black hole solutions
\cite{Dey:2004yt,Cai:2004eh,Guo:2022ghl}. The metric and electromagnetic
potential are given by
\begin{align}
ds^{2} &  =g_{\mu\nu}dx^{\mu}dx^{\nu}=-f_{\text{BI}}\left(  r\right)
dt^{2}+\frac{dr^{2}}{f_{\text{BI}}\left(  r\right)  }+r^{2}\left(  d\theta
^{2}+\sin^{2}\theta d\varphi^{2}\right)  \text{,}\nonumber\\
A_{\mu}dx^{\mu} &  =V_{BI}\left(  r\right)  dt,\label{eq:NLEDBH}%
\end{align}
where
\begin{align}
f_{\text{BI}}\left(  r\right)   &  =1-\frac{2M}{r}-\frac{2Q^{2}}{3\sqrt
{r^{4}+aQ^{2}}+3r^{2}}+\frac{4Q^{2}}{3r^{2}}\text{ }_{2}F_{1}\left(  \frac
{1}{4}\text{,}\frac{1}{2}\text{,}\frac{5}{4}\text{;}-\frac{aQ^{2}}{r^{4}%
}\right)  ,\nonumber\\
V_{BI}^{\prime}\left(  r\right)   &  =\frac{Q}{\sqrt{r^{4}+aQ^{2}}}.
\end{align}
Here, $M$ and $Q$ denote the black hole mass and electric charge,
respectively, and $_{2}F_{1}\left(  a,b,c;x\right)  $ is the hypergeometric
function. The solution exhibits a curvature singularity at $r=0$, as confirmed
by the calculation of the Kretschmann scalar \cite{Chen:2023trn},\qquad{}\
\begin{equation}
\mathcal{K}=R_{\mu\nu\rho\sigma}R^{\mu\nu\rho\sigma}\mathcal{=}\frac{16}{3\pi
r^{6}}\left[  3M\sqrt{\pi}-2a^{-1/4}Q^{4}\Gamma\left(  1/4\right)
\Gamma\left(  5/4\right)  \right]  ^{2}+\mathcal{O}\left(  r^{-5}\right)  .
\end{equation}
\begin{figure}[ptb]
\includegraphics[width=0.45\textwidth]{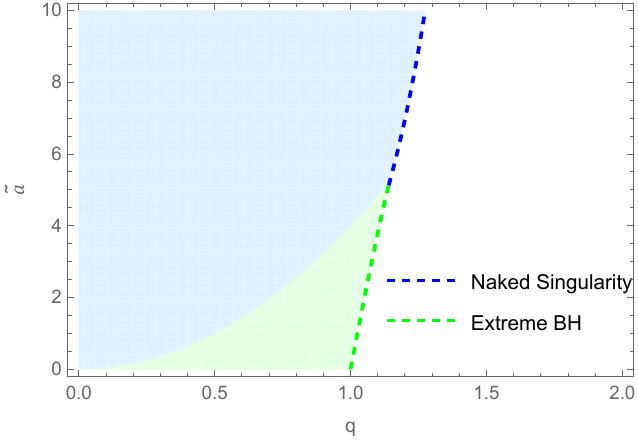} \hspace{15pt}
\includegraphics[width=0.45\textwidth]{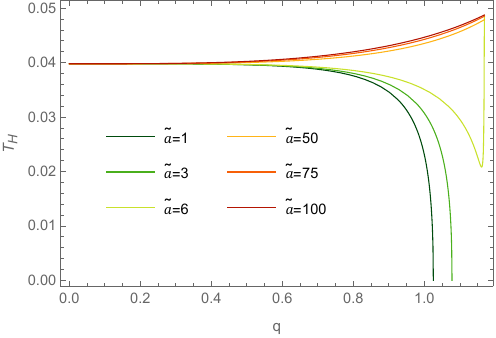}\caption{\textbf{Left Panel}:
Existence domain of static BI black holes in the $(q,\tilde{a})$ parameter
space. The blue region corresponds to black holes with a single horizon, while
the green region represents those with two horizons. The boundaries of the
colored regions are marked by dashed lines: the dashed blue line denotes naked
singularities, and the dashed green line corresponds to extremal black holes.
\textbf{Right Panel}: The Hawking temperature $T_{H}$ as a function of $q$ for
various values of $\tilde{a}$. For small $\tilde{a}$, $T_{H}$ approaches zero
at the extremal limit. For larger $\tilde{a}$, the temperature continues to
increase as the solution approaches a naked singularity.}%
\label{sBI}%
\end{figure}

Depending on the black hole parameters, static BI black holes can exhibit
either one or two horizons. The left panel of Fig. \ref{sBI} shows the domain
of existence in the $(q,\tilde{a})$ parameter space, where the dimensionless
quantities are defined as $q\equiv Q/M$ and $\tilde{a}\equiv a/M^{2}$. The
blue region corresponds to black holes with a single horizon, while the green
region indicates those with two horizons. The figure indicates that black hole
solutions cease to exist beyond certain critical values of $q$. At these
critical values, solutions with one horizon transition to naked singularities,
whereas those with two horizons approach extremal black holes. The right panel
displays the Hawking temperature $T_{H}$ as a function of $q$ for several
fixed values of $\tilde{a}$. For $\tilde{a}=1$ and $3$, the temperature
vanishes as $q$ approaches its critical value, indicating that the black holes
become extremal in this limit. In contrast, for $\tilde{a}=6$, $50$, $75$, and
$100$, the solutions at the critical $q$ correspond to naked singularities,
and $T_{H}$ exhibits\textbf{ }asymptotic growth as $q$ tends to the critical
point. It is worth noting that the non-monotonic behavior observed in the
$\tilde{a}=6$ case suggests its proximity to the transition between extremal
black holes and naked singularities.

\subsection{Rotating Black Hole Solutions}

To construct stationary, axisymmetric, and asymptotically flat black hole
solutions, we adopt the following general ansatz for the metric and gauge
field \cite{Herdeiro:2015gia,Herdeiro:2020wei,Guo:2023mda},
\begin{align}
ds^{2}  &  =-e^{2F_{0}}Ndt^{2}+e^{2F_{1}}\left(  \frac{dr^{2}}{N}+r^{2}%
d\theta^{2}\right)  +e^{2F_{2}}r^{2}\sin^{2}\theta\left(  d\varphi-\frac
{W}{r^{2}}dt\right)  ^{2},\nonumber\\
A_{\mu}dx^{\mu}  &  =\left(  A_{t}-A_{\varphi}\frac{W}{r^{2}}\sin^{2}%
\theta\right)  dt+A_{\varphi}\sin^{2}\theta d\varphi, \label{eq:ansatz}%
\end{align}
where $N=1-r_{H}/r$, with $r_{H}$ denoting the event horizon radius. The
functions $F_{0}$, $F_{1}$, $F_{2}$, $W$, $A_{t}$ and $A_{\varphi}$ are
regular and depend only on the coordinates $r$ and $\theta$. In a stationary
and axisymmetric spacetime, the surface gravity $\kappa$ and Hawking
temperature $T_{H}$ at the event horizon are given by
\begin{align}
\kappa &  =-\frac{1}{2}\left(  \nabla_{\mu}\xi_{\nu}\right)  \left(
\nabla^{\mu}\xi^{\nu}\right)  ,\nonumber\\
T_{H}  &  =\frac{\kappa}{2\pi}=\frac{1}{4\pi r_{H}}e^{F_{0}\left(
r_{H},\theta\right)  -F_{1}\left(  r_{H},\theta\right)  },
\end{align}
where $\xi=\partial_{t}+\Omega_{H}\partial_{\varphi}$ is the Killing vector
generating the event horizon, and $\Omega_{H}$ denotes the angular velocity at
the horizon. For a rotating BI black hole, the black hole entropy is given by
$S=A_{H}/4$, where the horizon area $A_{H}$ reads
\begin{equation}
A_{H}=2\pi r_{H}^{2}\int_{0}^{\pi}d\theta\sin\theta e^{F_{1}\left(
r_{H},\theta\right)  +F_{2}\left(  r_{H},\theta\right)  }.
\end{equation}

The asymptotic behavior of the metric and gauge field functions at the horizon
and at spatial infinity allows us to extract several physical quantities,
including the black hole mass $M$, black hole charge $Q$, magnetic dipole
moment $\mu_{M}$, black hole angular momentum $J$, electrostatic potential
$\Phi$, and horizon angular velocity $\Omega_{H}$
\cite{Herdeiro:2015gia,Delgado:2016jxq},
\begin{equation}
\left.  W\right\vert _{r=r_{H}}\sim r_{H}^{2}\Omega_{H},\text{ }\left.
W\right\vert _{r=\infty}\sim\frac{2J}{r},\text{ }\left.  e^{2F_{0}%
}N\right\vert _{r=\infty}\sim1-\frac{2M}{r},\text{ }\left.  A_{t}\right\vert
_{r=\infty}\sim\Phi-\frac{Q}{r},\quad A_{\varphi}|_{r=\infty}\sim\frac{\mu
_{M}}{r}. \label{eq:asymptotic behaviors}%
\end{equation}
These physical quantities satisfy the Smarr relation \cite{Rasheed:1997ns},
\begin{equation}
M=2T_{H}S+2\Omega_{H}J-\int_{\Sigma}dS_{\mu}\left(  2T_{\enskip\nu}^{\mu}%
\xi^{\nu}-T\xi^{\mu}\right)  , \label{eq:Smarr relation}%
\end{equation}
where $\Sigma$ is a spacelike hypersurface extending from the event horizon
out to spatial infinity. The Smarr relation serves as a consistency check for
estimating the accuracy of our numerically constructed black hole solutions.

Light rings and timelike circular geodesics play a fundamental role in black
hole physics. Light rings govern the formation of black hole shadows and
encode key information about the underlying spacetime geometry through their
connection to quasinormal modes. Meanwhile, the ISCOs determine the inner edge
of accretion disks and are critical for evaluating the efficiency of energy
extraction from black holes. In nonlinear electrodynamics, self-interactions
of the electromagnetic field modify photon propagation, such that photons
follow null geodesics in an effective geometry rather than in the background
black hole spacetime \cite{Novello:1999pg,He:2022opa}. A detailed
investigation of photon orbits in rotating BI black holes is beyond the scope
of this work and is left for future study. In this paper, we focus on the
properties of ISCOs confined to the equatorial plane.

Timelike circular geodesics in the equatorial plane satisfy the condition
\begin{equation}
g_{\mu\nu}\frac{dx^{\mu}}{d\lambda}\frac{dx^{\nu}}{d\lambda}=-e^{2F_{0}}%
N\dot{t}^{2}+e^{2F_{1}}\frac{\dot{r}^{2}}{N}+e^{2F_{2}}r^{2}\left(
\dot{\varphi}-\frac{W}{r^{2}}\dot{t}\right)  ^{2}=-1, \label{eq:circular geo}%
\end{equation}
where dots denote derivatives with respect to the affine parameter $\lambda$.
The geodesics admit two conserved quantities: the total energy $E$ and angular
momentum $L$ per unit mass, given by
\begin{equation}
E=\left(  e^{2F_{0}}N-e^{2F_{2}}\frac{W^{2}}{r^{2}}\right)  \dot{t}+e^{2F_{2}%
}W\dot{\varphi},\text{ }L=e^{2F_{2}}\left(  r^{2}\dot{\varphi}-W\dot
{t}\right)  . \label{eq:conservedq}%
\end{equation}
Substituting the expressions for $E$ and $L$ into Eq. $\left(
\ref{eq:circular geo}\right)  $, the geodesic equation reduces to a radial
equation of the form
\begin{equation}
\dot{r}^{2}+V_{\mathrm{eff}}=0,
\end{equation}
where the effective potential $V_{\mathrm{eff}}\left(  r\right)  $ is given
by
\begin{equation}
V_{\mathrm{eff}}=e^{-2F_{1}}N\left[  -\frac{e^{-2F_{0}}L^{2}}{N}\left(
\frac{E}{L}-H_{+}\right)  \left(  \frac{E}{L}-H_{-}\right)  +1\right]  ,
\end{equation}
with
\begin{equation}
H_{\pm}=\frac{W\pm\sqrt{e^{2F_{0}-2F_{2}}Nr^{2}}}{r^{2}}\text{.}%
\end{equation}
The ISCO radius $r_{\mathrm{ISCO}}$ is determined by
\begin{equation}
V_{\mathrm{eff}}\left(  r_{\mathrm{ISCO}}\right)  =0,\quad V_{\mathrm{eff}%
}^{\prime}\left(  r_{\mathrm{ISCO}}\right)  =0,\quad V_{\mathrm{eff}}%
^{\prime\prime}\left(  r_{\mathrm{ISCO}}\right)  =0. \label{eq:VISCO}%
\end{equation}
In spherically symmetric spacetimes, there typically exists a single ISCO.
However, in rotating black hole spacetimes, rotation causes a splitting
between prograde and retrograde ISCOs. The prograde ISCO generally has a
smaller radius and lower angular momentum. In contrast, the retrograde ISCO
occurs at a larger radius and requires greater angular momentum.

\subsection{Numerical Scheme}

We employ spectral methods to solve the coupled nonlinear partial differential
equations governing the system, obtained by substituting the ansatz $\left(
\ref{eq:ansatz}\right)  $ into the equations of motion $\left(
\ref{eq:NLEDEOM}\right)  $. For numerical implementation, we compactify the
radial coordinate $r$ via the transformation
\begin{equation}
x=\frac{\sqrt{r^{2}-r_{H}^{2}}-r_{H}}{\sqrt{r^{2}-r_{H}^{2}}+r_{H}},
\end{equation}
which maps the event horizon $r=r_{H}$ and spatial infinity $r=\infty$ to
$x=-1$ and $x=1$. Using this compactified coordinate $x$, the power series
expansions near the horizon yield the following boundary conditions at
$x=-1$:
\begin{equation}
\partial_{x}F_{0}=\partial_{x}F_{1}=\partial_{x}F_{2}=\partial_{x}A_{\varphi
}=A_{t}=W-\Omega_{H}=0. \label{eq:bdx0}%
\end{equation}
At spatial infinity $\left(  x=1\right)  $, the boundary conditions are
determined by the asymptotic behavior of the metric and gauge field
functions:
\begin{equation}
F_{0}=F_{1}=F_{2}=A_{\varphi}=2r_{H}\partial_{x}A_{t}-Q=-r_{H}\partial
_{x}W-\chi r_{H}^{2}\left(  \frac{1}{2}+2\partial_{x}F_{0}\right)  ^{2}=0,
\label{eq:bdx1}%
\end{equation}
where $\chi\equiv J/M^{2}$ denotes the dimensionless spin parameter. On the
symmetric axis $\theta=0$ and $\theta=\pi$, axial symmetry and regularity
impose the following conditions:
\begin{equation}
\partial_{\theta}F_{0}=\partial_{\theta}F_{1}=\partial_{\theta}F_{2}%
=\partial_{\theta}A_{\varphi}=\partial_{\theta}A_{t}=\partial_{\theta}W=0.
\label{eq:bdtheta0}%
\end{equation}

In this paper, we focus on solutions with equatorial-plane symmetry, allowing
the computational domain to be restricted to the upper half-plane $0\leq
\theta\leq\pi/2$. As a result, the boundary condition at $\theta=\pi$ in Eq.
$\left(  \ref{eq:bdtheta0}\right)  $ is replaced by
\begin{equation}
\partial_{\theta}F_{0}=\partial_{\theta}F_{1}=\partial_{\theta}F_{2}%
=\partial_{\theta}A_{\varphi}=\partial_{\theta}A_{t}=\partial_{\theta
}W=0\text{ at }\theta=\pi/2. \label{eq:bdtheta1}%
\end{equation}
Thus, Eqs. $\left(  \ref{eq:bdx0}\right)  $, $\left(  \ref{eq:bdx1}\right)  $,
$\left(  \ref{eq:bdtheta0}\right)  $ and $\left(  \ref{eq:bdtheta1}\right)  $
collectively define the boundary conditions used to solve the partial
differential equations. Furthermore, the absence of conical singularities on
the symmetry axis requires $F_{1}=F_{2}$, which provides an additional
consistency check for our numerical solutions, alongside the Smarr relation.

Spectral methods are a well-established and highly effective approach for
solving partial differential equations, particularly nonlinear elliptic
equations. These methods approximate solutions by expressing them as finite
linear combinations of basis functions, thereby transforming the differential
equations into a system of algebraic equations. As the resolution---i.e., the
number of basis functions---increases, the approximation exhibits exponential
convergence. This rapid convergence significantly outperforms the linear or
polynomial convergence rates typically associated with finite difference and
finite element methods.

In our numerical implementation, spectral methods are employed to approximate
the functions in the set $\mathcal{F}=\{F_{0},F_{1},F_{2},W,A_{t},A_{\varphi
}\}$ as a finite linear combination of basis functions,
\begin{equation}
\mathcal{F}^{\left(  k\right)  }=\sum_{i=0}^{N_{x}-1}\sum_{j=0}^{N_{\theta}%
-1}a_{ij}^{\left(  k\right)  }T_{i}\left(  x\right)  \cos\left(
2j\theta\right)  , \label{eq:F}%
\end{equation}
where $T_{i}\left(  x\right)  $ denotes the $i$-th Chebyshev polynomial,
$a_{ij}^{\left(  k\right)  }$ are the spectral coefficients, and $N_{x}$ and
$N_{\theta}$ represent the resolutions in the radial and angular directions,
respectively. By substituting the truncated series in Eq. $\left(
\ref{eq:F}\right)  $ into the coupled partial differential equations and
evaluating them at the Gauss-Chebyshev points, we obtain a system of algebraic
equations for the coefficients $a_{ij}^{\left(  k\right)  }$. This nonlinear
system is then solved using the Newton-Raphson method, where each root-finding
iteration employs Mathematica's built-in LinearSolve function.

\section{Numerical Results}

\label{sec:NR}

Since the boundary conditions in Eqs. $\left(  \ref{eq:bdx0}\right)  $,
$\left(  \ref{eq:bdx1}\right)  $, $\left(  \ref{eq:bdtheta0}\right)  $ and
$\left(  \ref{eq:bdtheta1}\right)  $ are determined by the parameters $r_{H}$,
$a$, $Q$, and $\chi$, the resulting black hole solutions are fully specified
by this set of parameters. In our numerical construction, we fix the horizon
radius $r_{H}$, so the solutions depend only on the dimensionless parameters
$a/r_{H}^{2}$, $Q/r_{H}$, and $\chi$. For fixed values of $a/r_{H}^{2}$ and
$\chi$, we construct a sequence of black hole solutions starting from
$Q/r_{H}=0$, gradually increasing the charge until numerical solutions can no
longer be obtained. This sequence is computed iteratively using the
Newton-Raphson method, with each solution obtained by taking the preceding one
with slightly smaller charge as the initial guess. To ensure numerical
accuracy and efficiency, we solve the partial differential equations using
spectral methods with resolutions $N_{x}=40$ and $N_{\theta}=10$.

\begin{figure}[ptb]
\includegraphics[width=0.32\textwidth]{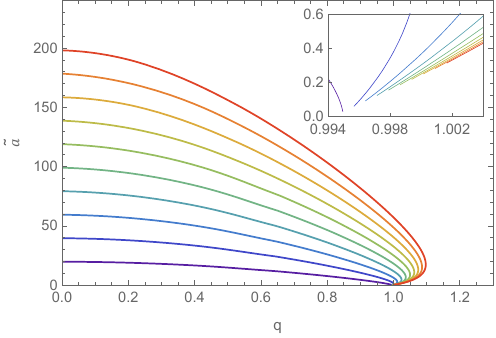} \hspace{2pt}
\includegraphics[width=0.32\textwidth]{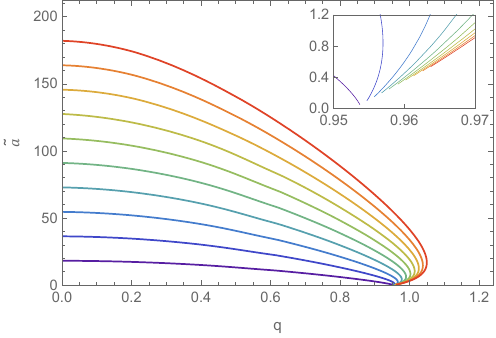} \hspace{2pt}
\includegraphics[width=0.32\textwidth]{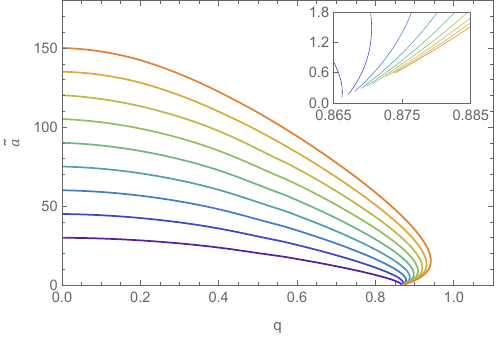}\caption{Constant-$a/r_{H}%
^{2}$ line in the $\left(  q,\tilde{a}\right)  $ parameter space for
$\chi=0.1$ (\textbf{Left}), $0.3$ (\textbf{Middle}), and $0.5$ (\textbf{Right}%
). From bottom to top, these lines correspond to $a/r_{H}^{2}=5$, $10$, $15$,
$20$, $\cdots$, $50$. Along each line, the charge-to-mass ratio $q$ increases
from zero to a maximum before decreasing. The lines terminate where numerical
solutions are no longer obtainable, indicating an approach to extremal black
hole solutions, as evidenced by the decreasing Hawking temperature. Insets
provide magnified views of the line endpoints, showing non-zero $\tilde{a}$
values that decrease with decreasing $a/r_{H}^{2}$.}%
\label{arh}%
\end{figure}

Fig. \ref{arh} shows lines of constant $a/r_{H}^{2}$ with $a/r_{H}^{2}=5$,
$10$, $15$, $20$, $\cdots$, $50$ in the $\left(  q,\tilde{a}\right)  $
parameter plane for $\chi=0.1$, $0.3$, and $0.5$. Along each line, the black
hole charge-to-mass ratio $q$ increases from zero to a maximum, and then
decreases. These constant-$a/r_{H}^{2}$ lines terminate when numerical
solutions can no longer be obtained within the prescribed tolerance. We find
that the Hawking temperature $T_{H}$ decreases monotonically as the endpoints
of these lines are approached, suggesting that the solutions approach the
extremal black hole limit. However, the numerical ansatz employed in this
paper does not yield sufficiently accurate black hole solutions in the
vicinity of the extremal limit. A detailed investigation of near-extremal BI
black holes is left for future work. The insets zoom in on the regions near
the endpoints, showing that the terminating values of $\tilde{a}$ remain
nonzero, and that these values decrease toward zero as $a/r_{H}^{2}$ is reduced.

In a more physically relevant scenario, we consider rotating BI black hole
solutions with a fixed value of $\tilde{a}$, which can be identified from the
intersections of the corresponding constant-$\tilde{a}$ line with various
constant-$a/r_{H}^{2}$ lines in the $\left(  q,\tilde{a}\right)  $ parameter
plane. As shown in Fig. \ref{arh}, for sufficiently small values of $\tilde
{a}$, the horizontal constant-$\tilde{a}$ line lies below the turning points
(i.e., the maxima of $q$) of certain constant-$a/r_{H}^{2}$ lines. As $q$
increases from zero, the constant-$\tilde{a}$ line may reach the endpoint of
one such line, suggesting that rotating BI black holes with small $\tilde{a}$
approach extremality as $q$ increases. In contrast, for sufficiently large
values of $\tilde{a}$, the constant-$\tilde{a}$ line lies above the turning
points of the constant-$a/r_{H}^{2}$ lines. As $q$ increases from zero, the
line does not reach any endpoint but continues to intersect constant-$a/r_{H}%
^{2}$ lines with increasingly large $a/r_{H}^{2}$ values. This behavior
implies that, for large $\tilde{a}$, rotating BI black holes approach naked
singularities rather than extremal black holes as $q$ increases.

\begin{figure}[ptb]
\includegraphics[width=0.32\textwidth]{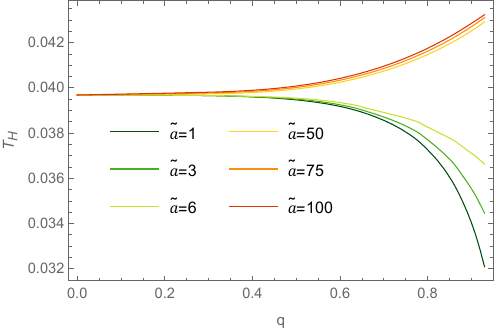} \hspace{2pt}
\includegraphics[width=0.32\textwidth]{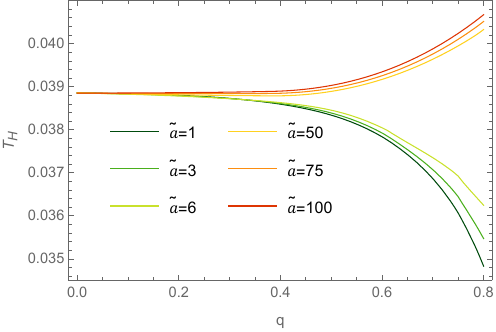} \hspace{2pt}
\includegraphics[width=0.32\textwidth]{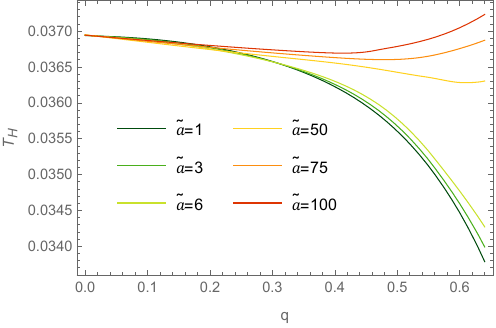}\caption{Hawking
temperatures $T_{H}$ versus the charge-to-mass ratio $q$ for rotating BI black
holes with spin parameters $\chi=0.1$ (\textbf{Left}), $0.3$ (\textbf{Middle}%
), and $0.5$ (\textbf{Right}). For small values of $\tilde{a}$ (e.g.,
$\tilde{a}=1$, $3$, and $6$), the temperature decreases monotonically with
increasing $q$, indicating that the black holes approach the extremal limit.
In contrast, for large $\tilde{a}$ (e.g., $\tilde{a}=50$, $75$, and $100$),
the temperature either increases monotonically or exhibits a non-monotonic
behavior, rising beyond a certain value of $q$, suggesting an approach to
naked singularities.}%
\label{fig:HT}%
\end{figure}

Fig. \ref{fig:HT} presents the Hawking temperatures of rotating BI black holes
as a function of $q$ for various values of $\tilde{a}$ at $\chi=0.1$, $0.3$,
and $0.5$. For $\tilde{a}=1$, $3$, and $6$, the Hawking temperature decreases
monotonically with increasing $q$, indicating that the black holes approach
extremal solutions, as expected. Furthermore, the results show that extremal
BI black holes occur at smaller values of $q$ as $\chi$ increases, reflecting
the influence of spin on the extremality condition. Interestingly, as shown in
Fig. \ref{sBI}, static BI black holes with $\tilde{a}=6$ approach naked
singularities as $q$ increases. This contrast suggests that rotation can shift
the critical behavior, favoring extremal black holes over naked singularities
at higher values of $\chi$. For larger values of $\tilde{a}=50$, $75$, and
$100$, the Hawking temperature behavior changes. In the $\chi=0.1$ and $0.3$
cases, $T_{H}$ increases monotonically with $q$, while in the $\chi=0.5$ case,
it initially decreases and then increases beyond a certain value of $q$. These
trends are consistent with the expectation that black holes with large
$\tilde{a}$ tend toward naked singularities rather than extremal solutions as
the charge increases.

\begin{figure}[ptb]
\includegraphics[width=0.32\textwidth]{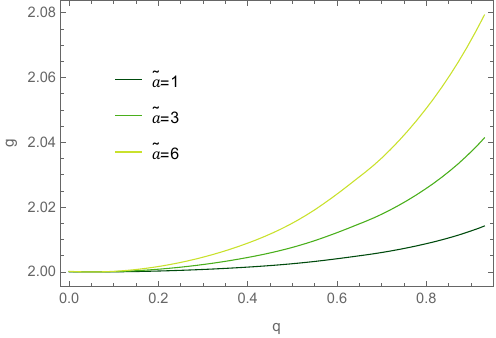} \hspace{2pt}
\includegraphics[width=0.32\textwidth]{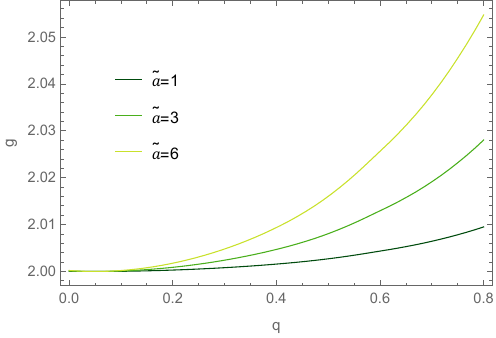} \hspace{2pt}
\includegraphics[width=0.32\textwidth]{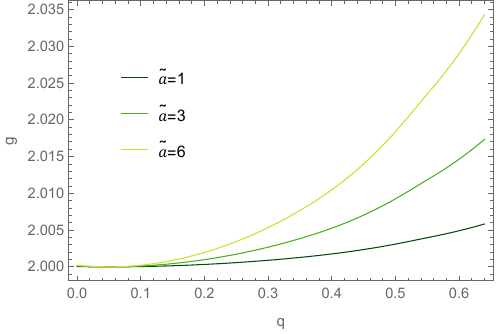}
\includegraphics[width=0.32\textwidth]{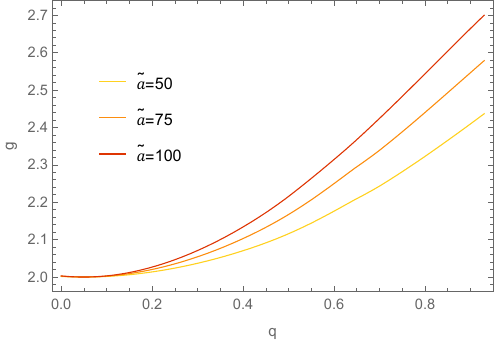} \hspace{2pt}
\includegraphics[width=0.32\textwidth]{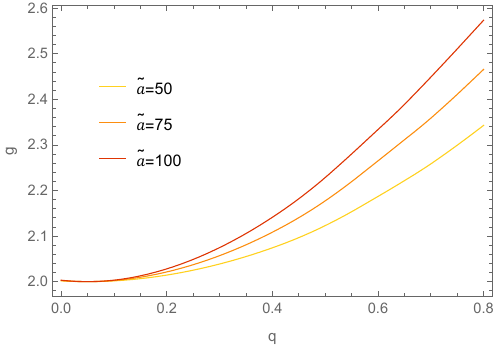} \hspace{2pt}
\includegraphics[width=0.32\textwidth]{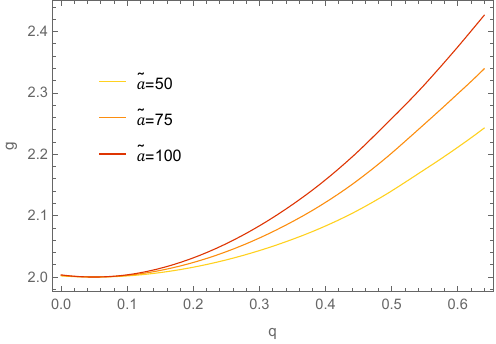}\caption{Gyromagnetic
ratio $g$ as a function of the charge-to-mass ratio $q$ for rotating BI black
holes with spin parameters $\chi=0.1$ (\textbf{Left}), $0.3$ (\textbf{Middle}%
), and $0.5$ (\textbf{Right}). The top and bottom rows correspond to
$\tilde{a}=1$, $3$, $6$ and $\tilde{a}=50$, $75$, $100$, respectively. In all
cases, the gyromagnetic ratio $g$ exceeds $2$ and increases with both $q$ and
$\tilde{a}$.}%
\label{fig:gfactor}%
\end{figure}

In rotating black hole spacetimes, electric charges induces a magnetic dipole
moment. Accordingly, the gyromagnetic ratio $g$ is defined as
\begin{equation}
g=\frac{2\mu_{M}M}{QJ},
\end{equation}
which characterizes the extent to which the magnetic dipole moment $\mu_{M}$
is generated by the black hole angular momentum $J$ and electric charge $Q$.
In Fig. \ref{fig:gfactor}, we present the gyromagnetic ratio $g$ of rotating
BI black holes as a function of $q$ for various values of $\tilde{a}$ at
$\chi=0.1$, $0.3$, and $0.5$, shown in the left, middle, and right columns,
respectively. In contrast to the KN black hole case, where $g=2$, our
numerical results reveal that $g$ is always greater than $2$ for rotating
charged BI black holes. The value of $g$ increases with both $q$ and the
dimensionless coupling parameter $\tilde{a}$, while exhibiting only a weak
dependence on the spin parameter $\chi$. Notably, for large $\tilde{a}$, the
deviation of $g$ from $2$ becomes significant. As expected, $g$ approaches $2$
in the limit $q\rightarrow0$, consistent with the diminishing effect of
nonlinear electrodynamics.

\begin{figure}[ptb]
\includegraphics[width=0.32\textwidth]{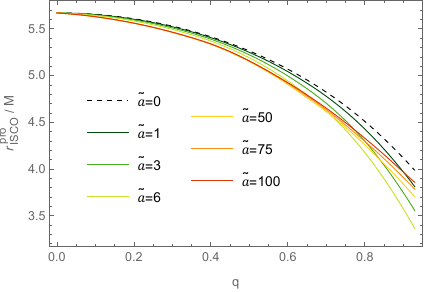} \hspace{2pt}
\includegraphics[width=0.32\textwidth]{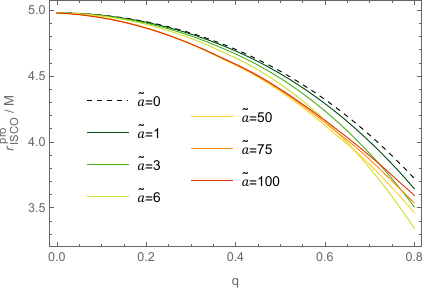} \hspace{2pt}
\includegraphics[width=0.32\textwidth]{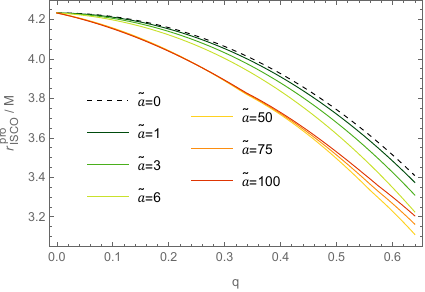}
\includegraphics[width=0.32\textwidth]{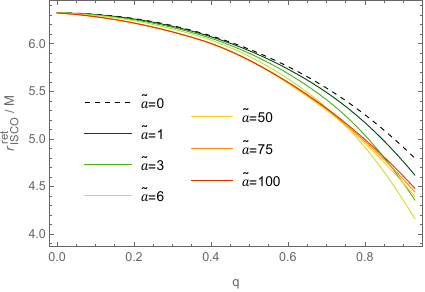} \hspace{2pt}
\includegraphics[width=0.32\textwidth]{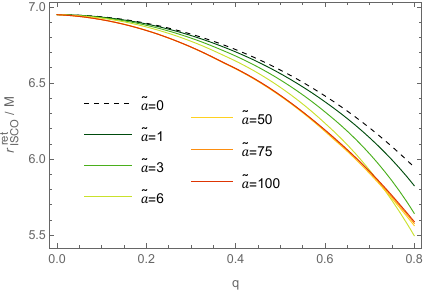} \hspace{2pt}
\includegraphics[width=0.32\textwidth]{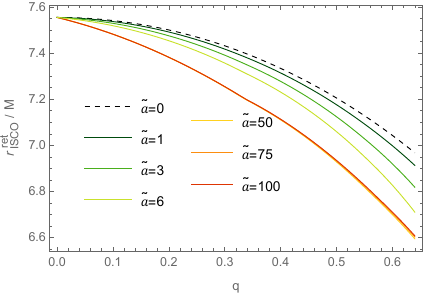}\caption{Prograde ISCO
radius $r_{\mathrm{ISCO}}^{\mathrm{pro}}/M$ (\textbf{Top}) and retrograde ISCO
radius $r_{\mathrm{ISCO}}^{\mathrm{ret}}/M$ (\textbf{Bottom}) as functions of
the charge-to-mass ratio $q$ for rotating BI black holes with spin parameters
$\chi=0.1$ (\textbf{Left}), $0.3$ (\textbf{Middle}), and $0.5$ (\textbf{Right}%
). Each panel shows results for various values $\tilde{a}$, along with KN
black holes ($\tilde{a}=0$, shown as dashed lines) for comparison. For all
values of $\tilde{a}$, both ISCO radii decrease monotonically with increasing
$q$. The dependence on $\tilde{a}$ is non-monotonic: ISCO radii decrease with
$\tilde{a}$ at small values of $\tilde{a}$, but increase with $\tilde{a}$ at
larger values. In all cases, rotating BI black holes exhibit smaller ISCO
radii than their Kerr-Newman counterparts.}%
\label{fig:isco}%
\end{figure}

Fig. \ref{fig:isco} displays the prograde ISCO radius $r_{\mathrm{ISCO}%
}^{\mathrm{pro}}$ and retrograde ISCO radius $r_{\mathrm{ISCO}}^{\mathrm{ret}%
}$ of rotating BI black holes as a function of $q$ for various values of
$\tilde{a}$. The results are presented for $\chi=0.1$, $0.3$, and $0.5$ in the
left, middle, and right columns, respectively. For comparison with KN black
holes, we also include the ISCO radii for $\tilde{a}=0$. Similar to the KN
case, both the prograde and retrograde ISCO radii decrease with increasing $q$
for all values of $\tilde{a}$. However, the dependence of the ISCO radii on
$\tilde{a}$ is more complex: for small values of $\tilde{a}$ (e.g., $\tilde
{a}=1$, $3$, and $6$), the ISCO radii decrease as $\tilde{a}$ increases,
whereas for large values of $\tilde{a}$ (e.g., $\tilde{a}=50$, $75$, and
$100$), the ISCO radii increase with $\tilde{a}$. Notably, the ISCO radii of
rotating BI black holes are consistently smaller than those of their KN counterparts.

\section{Conclusion}

\label{sec:Conclusion}

In this paper, we constructed and analyzed a family of rotating, charged black
hole solutions in EBI theory using spectral methods. We found that when
nonlinear electromagnetic effects are weak (i.e., small values of $\tilde{a}%
$), rotating BI black holes with fixed spin tend to approach extremality as
the electric charge increases. In contrast, for strong nonlinear effects
(i.e., large values of $\tilde{a}$), the solutions approach naked
singularities. This distinction is reflected in the behavior of the Hawking
temperature, which vanishes in the extremal limit but increases sharply near
the onset of naked singularities. We also investigated the electromagnetic
properties of rotating BI black holes, showing that their gyromagnetic ratio
$g$ consistently exceeds the KN value of $g=2$, with the deviation increasing
for larger values of charge and nonlinear coupling. Furthermore, we examined
the ISCOs and found that both prograde and retrograde ISCO radii are smaller
than those of KN black holes and exhibit a non-monotonic dependence on the
nonlinear coupling parameter.

Our study explores the properties of rotating BI black holes and lays the
groundwork for future investigations. The increase in gyromagnetic ratio and
the reduction in ISCO radius suggest potentially observable deviations from
the predictions of standard electrovacuum solutions. These effects may
influence accretion disk dynamics, electromagnetic emissions, and
gravitational wave signals in astrophysical black hole systems. Future
research directions include exploring the impact of nonlinear electrodynamics
on additional black hole observables, such as quasinormal mode spectra and
shadow radii, as well as extending the analysis to other models of nonlinear
electrodynamics. A more detailed study of the extremal limit and near-horizon
geometries would also be of significant interest.

\begin{acknowledgments}
We are grateful to Guangzhou Guo and Tianshu Wu for useful discussions and
valuable comments. This work is supported in part by NSFC (Grant Nos. 12275183
and 12275184).
\end{acknowledgments}

\bibliographystyle{unsrturl}
\bibliography{ref}

\begin{thebibliography}{10}

\bibitem{Newman:1965my}
E~T. Newman, E.~Couch, K.~Chinnapared, A.~Exton, A.~Prakash, and R.~Torrence.
\newblock {Metric of a Rotating, Charged Mass}.
\newblock {\em J. Math. Phys.}, 6:918--919, 1965.
\newblock \href {https://doi.org/10.1063/1.1704351}
  {\path{doi:10.1063/1.1704351}}.

\bibitem{Soleng:1995kn}
Harald~H. Soleng.
\newblock {Charged black points in general relativity coupled to the
  logarithmic U(1) gauge theory}.
\newblock {\em Phys. Rev. D}, 52:6178--6181, 1995.
\newblock \href {https://arxiv.org/abs/hep-th/9509033}
  {\path{arXiv:hep-th/9509033}}, \href
  {https://doi.org/10.1103/PhysRevD.52.6178}
  {\path{doi:10.1103/PhysRevD.52.6178}}.

\bibitem{AyonBeato:1998ub}
Eloy Ayon-Beato and Alberto Garcia.
\newblock {Regular black hole in general relativity coupled to nonlinear
  electrodynamics}.
\newblock {\em Phys. Rev. Lett.}, 80:5056--5059, 1998.
\newblock \href {https://arxiv.org/abs/gr-qc/9911046}
  {\path{arXiv:gr-qc/9911046}}, \href
  {https://doi.org/10.1103/PhysRevLett.80.5056}
  {\path{doi:10.1103/PhysRevLett.80.5056}}.

\bibitem{Born:1934gh}
M.~Born and L.~Infeld.
\newblock {Foundations of the new field theory}.
\newblock {\em Proc. Roy. Soc. Lond. A}, 144(852):425--451, 1934.
\newblock \href {https://doi.org/10.1098/rspa.1934.0059}
  {\path{doi:10.1098/rspa.1934.0059}}.

\bibitem{Fradkin:1985qd}
E.~S. Fradkin and Arkady~A. Tseytlin.
\newblock {Nonlinear Electrodynamics from Quantized Strings}.
\newblock {\em Phys. Lett. B}, 163:123--130, 1985.
\newblock \href {https://doi.org/10.1016/0370-2693(85)90205-9}
  {\path{doi:10.1016/0370-2693(85)90205-9}}.

\bibitem{Tseytlin:1986ti}
Arkady~A. Tseytlin.
\newblock {Vector Field Effective Action in the Open Superstring Theory}.
\newblock {\em Nucl. Phys. B}, 276:391, 1986.
\newblock [Erratum: Nucl.Phys.B 291, 876 (1987)].
\newblock \href {https://doi.org/10.1016/0550-3213(86)90303-2}
  {\path{doi:10.1016/0550-3213(86)90303-2}}.

\bibitem{Salazar:1987ap}
I.~H. Salazar, A.~Garcia, and J.~Plebanski.
\newblock {Duality Rotations and Type $D$ Solutions to Einstein Equations With
  Nonlinear Electromagnetic Sources}.
\newblock {\em J. Math. Phys.}, 28:2171--2181, 1987.
\newblock \href {https://doi.org/10.1063/1.527430}
  {\path{doi:10.1063/1.527430}}.

\bibitem{Wiltshire:1988uq}
David~L. Wiltshire.
\newblock {Black Holes in String Generated Gravity Models}.
\newblock {\em Phys. Rev. D}, 38:2445, 1988.
\newblock \href {https://doi.org/10.1103/PhysRevD.38.2445}
  {\path{doi:10.1103/PhysRevD.38.2445}}.

\bibitem{Demianski:1986wx}
M.~Demianski.
\newblock {STATIC ELECTROMAGNETIC GEON}.
\newblock {\em Found. Phys.}, 16:187--190, 1986.
\newblock \href {https://doi.org/10.1007/BF01889380}
  {\path{doi:10.1007/BF01889380}}.

\bibitem{Fernando:2003tz}
Sharmanthie Fernando and Don Krug.
\newblock {Charged black hole solutions in Einstein-Born-Infeld gravity with a
  cosmological constant}.
\newblock {\em Gen. Rel. Grav.}, 35:129--137, 2003.
\newblock \href {https://arxiv.org/abs/hep-th/0306120}
  {\path{arXiv:hep-th/0306120}}, \href
  {https://doi.org/10.1023/A:1021315214180}
  {\path{doi:10.1023/A:1021315214180}}.

\bibitem{Dey:2004yt}
Tanay~Kr. Dey.
\newblock {Born-Infeld black holes in the presence of a cosmological constant}.
\newblock {\em Phys. Lett. B}, 595(1-4):484--490, 2004.
\newblock \href {https://arxiv.org/abs/hep-th/0406169}
  {\path{arXiv:hep-th/0406169}}, \href
  {https://doi.org/10.1016/j.physletb.2004.06.047}
  {\path{doi:10.1016/j.physletb.2004.06.047}}.

\bibitem{Cai:2004eh}
Rong-Gen Cai, Da-Wei Pang, and Anzhong Wang.
\newblock {Born-Infeld black holes in (A)dS spaces}.
\newblock {\em Phys. Rev. D}, 70:124034, 2004.
\newblock \href {https://arxiv.org/abs/hep-th/0410158}
  {\path{arXiv:hep-th/0410158}}, \href
  {https://doi.org/10.1103/PhysRevD.70.124034}
  {\path{doi:10.1103/PhysRevD.70.124034}}.

\bibitem{Rasheed:1997ns}
D.~A. Rasheed.
\newblock {Nonlinear electrodynamics: Zeroth and first laws of black hole
  mechanics}.
\newblock 2 1997.
\newblock \href {https://arxiv.org/abs/hep-th/9702087}
  {\path{arXiv:hep-th/9702087}}.

\bibitem{Breton:2004qa}
Nora Breton.
\newblock {Smarr's formula for black holes with non-linear electrodynamics}.
\newblock {\em Gen. Rel. Grav.}, 37:643--650, 2005.
\newblock \href {https://arxiv.org/abs/gr-qc/0405116}
  {\path{arXiv:gr-qc/0405116}}, \href
  {https://doi.org/10.1007/s10714-005-0051-x}
  {\path{doi:10.1007/s10714-005-0051-x}}.

\bibitem{Banerjee:2010da}
Rabin Banerjee, Sumit Ghosh, and Dibakar Roychowdhury.
\newblock {New type of phase transition in Reissner Nordström--AdS black hole
  and its thermodynamic geometry}.
\newblock {\em Phys. Lett. B}, 696:156--162, 2011.
\newblock \href {https://arxiv.org/abs/1008.2644} {\path{arXiv:1008.2644}},
  \href {https://doi.org/10.1016/j.physletb.2010.12.010}
  {\path{doi:10.1016/j.physletb.2010.12.010}}.

\bibitem{Zou:2013owa}
De-Cheng Zou, Shao-Jun Zhang, and Bin Wang.
\newblock {Critical behavior of Born-Infeld AdS black holes in the extended
  phase space thermodynamics}.
\newblock {\em Phys. Rev. D}, 89(4):044002, Feb 2014.
\newblock URL: \url{https://link.aps.org/doi/10.1103/PhysRevD.89.044002}, \href
  {https://arxiv.org/abs/1311.7299} {\path{arXiv:1311.7299}}, \href
  {https://doi.org/10.1103/PhysRevD.89.044002}
  {\path{doi:10.1103/PhysRevD.89.044002}}.

\bibitem{Hendi:2015hoa}
Seyed~Hossein Hendi, Behzad Eslam~Panah, and Shahram Panahiyan.
\newblock {Einstein-Born-Infeld-Massive Gravity: adS-Black Hole Solutions and
  their Thermodynamical properties}.
\newblock {\em JHEP}, 11:157, 2015.
\newblock \href {https://arxiv.org/abs/1508.01311} {\path{arXiv:1508.01311}},
  \href {https://doi.org/10.1007/JHEP11(2015)157}
  {\path{doi:10.1007/JHEP11(2015)157}}.

\bibitem{Wang:2019kxp}
Peng Wang, Houwen Wu, and Haitang Yang.
\newblock {Thermodynamics and Phase Transition of a Nonlinear Electrodynamics
  Black Hole in a Cavity}.
\newblock {\em JHEP}, 07:002, 2019.
\newblock \href {https://arxiv.org/abs/1901.06216} {\path{arXiv:1901.06216}},
  \href {https://doi.org/10.1007/JHEP07(2019)002}
  {\path{doi:10.1007/JHEP07(2019)002}}.

\bibitem{Liang:2019dni}
Kangkai Liang, Peng Wang, Houwen Wu, and Mingtao Yang.
\newblock {Phase structures and transitions of Born--Infeld black holes in a
  grand canonical ensemble}.
\newblock {\em Eur. Phys. J. C}, 80(3):187, 2020.
\newblock \href {https://arxiv.org/abs/1907.00799} {\path{arXiv:1907.00799}},
  \href {https://doi.org/10.1140/epjc/s10052-020-7750-z}
  {\path{doi:10.1140/epjc/s10052-020-7750-z}}.

\bibitem{Wang:2018xdz}
Peng Wang, Houwen Wu, and Haitang Yang.
\newblock {Thermodynamics and Phase Transitions of Nonlinear Electrodynamics
  Black Holes in an Extended Phase Space}.
\newblock {\em JCAP}, 04(04):052, 2019.
\newblock \href {https://arxiv.org/abs/1808.04506} {\path{arXiv:1808.04506}},
  \href {https://doi.org/10.1088/1475-7516/2019/04/052}
  {\path{doi:10.1088/1475-7516/2019/04/052}}.

\bibitem{Breton:2017hwe}
Nora Bret{\'o}n, Tyler Clark, and Sharmanthie Fernando.
\newblock {Quasinormal modes and absorption cross-sections of
  Born{\textendash}Infeld{\textendash}de Sitter black holes}.
\newblock {\em Int. J. Mod. Phys. D}, 26(10):1750112, 2017.
\newblock \href {https://arxiv.org/abs/1703.10070} {\path{arXiv:1703.10070}},
  \href {https://doi.org/10.1142/S0218271817501127}
  {\path{doi:10.1142/S0218271817501127}}.

\bibitem{Lee:2020iau}
Chong~Oh Lee, Jin~Young Kim, and Mu-In Park.
\newblock {Quasi-normal modes and stability of
  Einstein{\textendash}Born{\textendash}Infeld black holes in de Sitter space}.
\newblock {\em Eur. Phys. J. C}, 80(8):763, 2020.
\newblock \href {https://arxiv.org/abs/2004.12185} {\path{arXiv:2004.12185}},
  \href {https://doi.org/10.1140/epjc/s10052-020-8309-8}
  {\path{doi:10.1140/epjc/s10052-020-8309-8}}.

\bibitem{He:2022opa}
Aoyun He, Jun Tao, Peng Wang, Yadong Xue, and Lingkai Zhang.
\newblock {Effects of Born\textendash{}Infeld electrodynamics on black hole
  shadows}.
\newblock {\em Eur. Phys. J. C}, 82(8):683, 2022.
\newblock \href {https://arxiv.org/abs/2205.12779} {\path{arXiv:2205.12779}},
  \href {https://doi.org/10.1140/epjc/s10052-022-10637-x}
  {\path{doi:10.1140/epjc/s10052-022-10637-x}}.

\bibitem{Chen:2023trn}
Yiqian Chen, Peng Wang, Houwen Wu, and Haitang Yang.
\newblock {Gravitational lensing by Born-Infeld naked singularities}.
\newblock {\em Phys. Rev. D}, 109(8):084014, 2024.
\newblock \href {https://arxiv.org/abs/2305.17411} {\path{arXiv:2305.17411}},
  \href {https://doi.org/10.1103/PhysRevD.109.084014}
  {\path{doi:10.1103/PhysRevD.109.084014}}.

\bibitem{Chen:2023knf}
Yiqian Chen, Peng Wang, Houwen Wu, and Haitang Yang.
\newblock {Observations of orbiting hot spots around naked singularities}.
\newblock {\em JCAP}, 04:032, 2024.
\newblock \href {https://arxiv.org/abs/2309.04157} {\path{arXiv:2309.04157}},
  \href {https://doi.org/10.1088/1475-7516/2024/04/032}
  {\path{doi:10.1088/1475-7516/2024/04/032}}.

\bibitem{Tao:2017fsy}
Jun Tao, Peng Wang, and Haitang Yang.
\newblock {Testing holographic conjectures of complexity with Born--Infeld
  black holes}.
\newblock {\em Eur. Phys. J. C}, 77(12):817, 2017.
\newblock \href {https://arxiv.org/abs/1703.06297} {\path{arXiv:1703.06297}},
  \href {https://doi.org/10.1140/epjc/s10052-017-5395-3}
  {\path{doi:10.1140/epjc/s10052-017-5395-3}}.

\bibitem{Gan:2019jac}
Qingyu Gan, Guangzhou Guo, Peng Wang, and Houwen Wu.
\newblock {Strong cosmic censorship for a scalar field in a
  Born-Infeld\textendash{}de Sitter black hole}.
\newblock {\em Phys. Rev. D}, 100(12):124009, 2019.
\newblock \href {https://arxiv.org/abs/1907.04466} {\path{arXiv:1907.04466}},
  \href {https://doi.org/10.1103/PhysRevD.100.124009}
  {\path{doi:10.1103/PhysRevD.100.124009}}.

\bibitem{Peng:2018smy}
Yan Peng.
\newblock {Hair distributions in noncommutative Einstein-Born-Infeld black
  holes}.
\newblock {\em Nucl. Phys. B}, 941:1--10, 2019.
\newblock \href {https://arxiv.org/abs/1808.07988} {\path{arXiv:1808.07988}},
  \href {https://doi.org/10.1016/j.nuclphysb.2019.02.016}
  {\path{doi:10.1016/j.nuclphysb.2019.02.016}}.

\bibitem{Wang:2020ohb}
Peng Wang, Houwen Wu, and Haitang Yang.
\newblock {Scalarized Einstein-Born-Infeld black holes}.
\newblock {\em Phys. Rev. D}, 103(10):104012, 2021.
\newblock \href {https://arxiv.org/abs/2012.01066} {\path{arXiv:2012.01066}},
  \href {https://doi.org/10.1103/PhysRevD.103.104012}
  {\path{doi:10.1103/PhysRevD.103.104012}}.

\bibitem{Nomura:2020tpc}
Kimihiro Nomura, Daisuke Yoshida, and Jiro Soda.
\newblock {Stability of magnetic black holes in general nonlinear
  electrodynamics}.
\newblock {\em Phys. Rev. D}, 101(12):124026, 2020.
\newblock \href {https://arxiv.org/abs/2004.07560} {\path{arXiv:2004.07560}},
  \href {https://doi.org/10.1103/PhysRevD.101.124026}
  {\path{doi:10.1103/PhysRevD.101.124026}}.

\bibitem{Yang:2020jno}
Ke~Yang, Bao-Min Gu, Shao-Wen Wei, and Yu-Xiao Liu.
\newblock {Born{\textendash}Infeld black holes in 4D
  Einstein{\textendash}Gauss{\textendash}Bonnet gravity}.
\newblock {\em Eur. Phys. J. C}, 80(7):662, 2020.
\newblock \href {https://arxiv.org/abs/2004.14468} {\path{arXiv:2004.14468}},
  \href {https://doi.org/10.1140/epjc/s10052-020-8246-6}
  {\path{doi:10.1140/epjc/s10052-020-8246-6}}.

\bibitem{Falciano:2021kdu}
F.~T. Falciano, M.~L. Pe{\~n}afiel, and J.~C. Fabris.
\newblock {Entropy bound in Einstein-Born-Infeld black holes}.
\newblock {\em Phys. Rev. D}, 103(8):084046, 2021.
\newblock \href {https://arxiv.org/abs/2103.14109} {\path{arXiv:2103.14109}},
  \href {https://doi.org/10.1103/PhysRevD.103.084046}
  {\path{doi:10.1103/PhysRevD.103.084046}}.

\bibitem{Babar:2021nst}
Gulmina~Zaman Babar, Farruh Atamurotov, Shafqat Ul~Islam, and Sushant~G. Ghosh.
\newblock {Particle acceleration around rotating Einstein-Born-Infeld black
  hole and plasma effect on gravitational lensing}.
\newblock {\em Phys. Rev. D}, 103(8):084057, 2021.
\newblock \href {https://arxiv.org/abs/2104.00714} {\path{arXiv:2104.00714}},
  \href {https://doi.org/10.1103/PhysRevD.103.084057}
  {\path{doi:10.1103/PhysRevD.103.084057}}.

\bibitem{Yang:2022qoc}
Yisong Yang.
\newblock {Dyonically charged black holes arising in generalized
  Born{\textendash}Infeld theory of electromagnetism}.
\newblock {\em Annals Phys.}, 443:168996, 2022.
\newblock \href {https://arxiv.org/abs/2204.11313} {\path{arXiv:2204.11313}},
  \href {https://doi.org/10.1016/j.aop.2022.168996}
  {\path{doi:10.1016/j.aop.2022.168996}}.

\bibitem{Ali:2023jox}
Md~Sabir Ali, Hasan El~Moumni, Jamal Khalloufi, and Karima Masmar.
\newblock {Topology of Born{\textendash}Infeld-AdS black hole phase
  transitions: Bulk and CFT sides}.
\newblock {\em Annals Phys.}, 465:169679, 2024.
\newblock \href {https://arxiv.org/abs/2306.11212} {\path{arXiv:2306.11212}},
  \href {https://doi.org/10.1016/j.aop.2024.169679}
  {\path{doi:10.1016/j.aop.2024.169679}}.

\bibitem{Tang:2023lmr}
Jiuyang Tang, Yunqi Liu, Wei-Liang Qian, and Ruihong Yue.
\newblock {Effect of nonlinear electrodynamics on shadows of slowly rotating
  black holes}.
\newblock {\em Chin. Phys. C}, 47(2):025105, 2023.
\newblock \href {https://doi.org/10.1088/1674-1137/ac9fba}
  {\path{doi:10.1088/1674-1137/ac9fba}}.

\bibitem{Wu:2024fvy}
Zhe-Hua Wu and H.~Lu.
\newblock {Superradiant instability of charged extremal black holes in
  Einstein-Born-Infeld gravity}.
\newblock {\em JHEP}, 07:003, 2024.
\newblock \href {https://arxiv.org/abs/2404.02977} {\path{arXiv:2404.02977}},
  \href {https://doi.org/10.1007/JHEP07(2024)003}
  {\path{doi:10.1007/JHEP07(2024)003}}.

\bibitem{Maier:2024ahy}
Rodrigo Maier and Manuella Corr{\^e}a~e Silva.
\newblock {Charged black holes from an interacting vacuum}.
\newblock {\em Phys. Rev. D}, 110(8):084079, 2024.
\newblock \href {https://arxiv.org/abs/2405.10293} {\path{arXiv:2405.10293}},
  \href {https://doi.org/10.1103/PhysRevD.110.084079}
  {\path{doi:10.1103/PhysRevD.110.084079}}.

\bibitem{Zahid:2024nvx}
Muhammad Zahid, Furkat Sarikulov, Chao Shen, Maksud Umaraliyev, and Javlon
  Rayimbaev.
\newblock {Shadow and quasinormal modes of novel charged rotating black hole in
  Born{\textendash}Infeld theory: Constraints from EHT results}.
\newblock {\em Phys. Dark Univ.}, 46:101616, 2024.
\newblock \href {https://doi.org/10.1016/j.dark.2024.101616}
  {\path{doi:10.1016/j.dark.2024.101616}}.

\bibitem{Wu:2024qpf}
Ke-Tai Wu, Zi-Jun Zhong, Yi~Li, Chong-Ye Chen, Cheng-Yong Zhang, Chao Niu, and
  Peng Liu.
\newblock {Dynamics of spontaneous scalarization of black holes with nonlinear
  electromagnetic fields in anti-de Sitter spacetime}.
\newblock 12 2024.
\newblock \href {https://arxiv.org/abs/2412.02132} {\path{arXiv:2412.02132}}.

\bibitem{Ye:2025xvi}
Guang-Zai Ye, Chong-Ye Chen, and Peng Liu.
\newblock {Spontaneous Vectorization in the Einstein-Born-Infeld-Vector Model}.
\newblock 4 2025.
\newblock \href {https://arxiv.org/abs/2504.09821} {\path{arXiv:2504.09821}}.

\bibitem{Chen:2025cwi}
Yiqian Chen, Guangzhou Guo, Benrong Mu, and Peng Wang.
\newblock {Schwarzschild Black Holes Immersed in Born-Infeld Magnetic Fields
  and Their Observational Signatures}.
\newblock 6 2025.
\newblock \href {https://arxiv.org/abs/2506.19581} {\path{arXiv:2506.19581}}.

\bibitem{CiriloLombardo:2004qw}
Diego~Julio Cirilo~Lombardo.
\newblock {The Newman-Janis algorithm, rotating solutions and
  Einstein-Born-Infeld black holes}.
\newblock {\em Class. Quant. Grav.}, 21:1407--1417, 2004.
\newblock \href {https://arxiv.org/abs/gr-qc/0612063}
  {\path{arXiv:gr-qc/0612063}}, \href
  {https://doi.org/10.1088/0264-9381/21/6/009}
  {\path{doi:10.1088/0264-9381/21/6/009}}.

\bibitem{CiriloLombardo:2005qmc}
Diego~Julio Cirilo~Lombardo.
\newblock {Rotating charged black holes in Einstein-Born-Infeld theories and
  their ADM mass}.
\newblock {\em Gen. Rel. Grav.}, 37:847--856, 2005.
\newblock \href {https://arxiv.org/abs/gr-qc/0603066}
  {\path{arXiv:gr-qc/0603066}}, \href
  {https://doi.org/10.1007/s10714-005-0071-6}
  {\path{doi:10.1007/s10714-005-0071-6}}.

\bibitem{Carter:1968rr}
Brandon Carter.
\newblock {Global structure of the Kerr family of gravitational fields}.
\newblock {\em Phys. Rev.}, 174:1559--1571, 1968.
\newblock \href {https://doi.org/10.1103/PhysRev.174.1559}
  {\path{doi:10.1103/PhysRev.174.1559}}.

\bibitem{Garfinkle:1990ib}
David Garfinkle and Jennie~H. Traschen.
\newblock {On the Gyromagnetic Ratio of a Black Hole}.
\newblock {\em Phys. Rev. D}, 42:419--423, 1990.
\newblock \href {https://doi.org/10.1103/PhysRevD.42.419}
  {\path{doi:10.1103/PhysRevD.42.419}}.

\bibitem{Aliev:2004ec}
A.~N. Aliev and Valeri~P. Frolov.
\newblock {Five-dimensional rotating black hole in a uniform magnetic field:
  The Gyromagnetic ratio}.
\newblock {\em Phys. Rev. D}, 69:084022, 2004.
\newblock \href {https://arxiv.org/abs/hep-th/0401095}
  {\path{arXiv:hep-th/0401095}}, \href
  {https://doi.org/10.1103/PhysRevD.69.084022}
  {\path{doi:10.1103/PhysRevD.69.084022}}.

\bibitem{Ortaggio:2006ng}
Marcello Ortaggio and Vojtech Pravda.
\newblock {Black rings with a small electric charge: Gyromagnetic ratios and
  algebraic alignment}.
\newblock {\em JHEP}, 12:054, 2006.
\newblock \href {https://arxiv.org/abs/gr-qc/0609049}
  {\path{arXiv:gr-qc/0609049}}, \href
  {https://doi.org/10.1088/1126-6708/2006/12/054}
  {\path{doi:10.1088/1126-6708/2006/12/054}}.

\bibitem{Aliev:2006tt}
A.~N. Aliev.
\newblock {Gyromagnetic Ratio of Charged Kerr-Anti-de Sitter Black Holes}.
\newblock {\em Class. Quant. Grav.}, 24:4669--4678, 2007.
\newblock \href {https://arxiv.org/abs/hep-th/0611205}
  {\path{arXiv:hep-th/0611205}}, \href
  {https://doi.org/10.1088/0264-9381/24/18/008}
  {\path{doi:10.1088/0264-9381/24/18/008}}.

\bibitem{Delgado:2016jxq}
Jorge F.~M. Delgado, Carlos A.~R. Herdeiro, Eugen Radu, and Helgi Runarsson.
\newblock {Kerr\textendash{}Newman black holes with scalar hair}.
\newblock {\em Phys. Lett. B}, 761:234--241, 2016.
\newblock \href {https://arxiv.org/abs/1608.00631} {\path{arXiv:1608.00631}},
  \href {https://doi.org/10.1016/j.physletb.2016.08.032}
  {\path{doi:10.1016/j.physletb.2016.08.032}}.

\bibitem{Guo:2022ghl}
Guangzhou Guo, Yuhang Lu, Peng Wang, Houwen Wu, and Haitang Yang.
\newblock {Black holes with multiple photon spheres}.
\newblock {\em Phys. Rev. D}, 107(12):124037, 2023.
\newblock \href {https://arxiv.org/abs/2212.12901} {\path{arXiv:2212.12901}},
  \href {https://doi.org/10.1103/PhysRevD.107.124037}
  {\path{doi:10.1103/PhysRevD.107.124037}}.

\bibitem{Herdeiro:2015gia}
Carlos Herdeiro and Eugen Radu.
\newblock {Construction and physical properties of Kerr black holes with scalar
  hair}.
\newblock {\em Class. Quant. Grav.}, 32(14):144001, 2015.
\newblock \href {https://arxiv.org/abs/1501.04319} {\path{arXiv:1501.04319}},
  \href {https://doi.org/10.1088/0264-9381/32/14/144001}
  {\path{doi:10.1088/0264-9381/32/14/144001}}.

\bibitem{Herdeiro:2020wei}
Carlos A.~R. Herdeiro, Eugen Radu, Hector~O. Silva, Thomas~P. Sotiriou, and
  Nicol\'as Yunes.
\newblock {Spin-induced scalarized black holes}.
\newblock {\em Phys. Rev. Lett.}, 126(1):011103, 2021.
\newblock \href {https://arxiv.org/abs/2009.03904} {\path{arXiv:2009.03904}},
  \href {https://doi.org/10.1103/PhysRevLett.126.011103}
  {\path{doi:10.1103/PhysRevLett.126.011103}}.

\bibitem{Guo:2023mda}
Guangzhou Guo, Peng Wang, Houwen Wu, and Haitang Yang.
\newblock {Scalarized Kerr-Newman black holes}.
\newblock {\em JHEP}, 10:076, 2023.
\newblock \href {https://arxiv.org/abs/2307.12210} {\path{arXiv:2307.12210}},
  \href {https://doi.org/10.1007/JHEP10(2023)076}
  {\path{doi:10.1007/JHEP10(2023)076}}.

\bibitem{Novello:1999pg}
M.~Novello, V.~A. De~Lorenci, J.~M. Salim, and Renato Klippert.
\newblock {Geometrical aspects of light propagation in nonlinear
  electrodynamics}.
\newblock {\em Phys. Rev. D}, 61:045001, 2000.
\newblock \href {https://arxiv.org/abs/gr-qc/9911085}
  {\path{arXiv:gr-qc/9911085}}, \href
  {https://doi.org/10.1103/PhysRevD.61.045001}
  {\path{doi:10.1103/PhysRevD.61.045001}}.

\end{thebibliography}

\end{document}